\documentclass{PoS}

\title{Multiple Critical Points in the QCD Phase Diagram}

\ShortTitle{Multiple Critical Points}

\DeclareSymbolFont{letters}{OML}{cmm}{m}{it}
\SetSymbolFont{letters}{bold}{OML}{cmm}{b}{it}
\usepackage{amsmath}

\author{\speaker{J. I. Kapusta} and E. S. Bowman\\
School of Physics and Astronomy, University of Minnesota,
Minneapolis, MN, 55455, USA \\
        E-mail: \email{kapusta@physics.umn.edu}}

\abstract{We use the linear $\sigma$ model with two flavors of quarks to study the phase diagram at finite temperature and baryon chemical potential as a function of the vacuum pion mass.  Our calculations include thermal fluctuations of both the bosonic and fermionic fields.  An interesting phase structure occurs that results in two, one, or no critical points depending on the value of the pion mass.}

\FullConference{5th International Workshop on Critical Point and Onset of Deconfinement - CPOD 2009,\\
		 June 08 - 12 2009\\
		 Brookhaven National Laboratory, Long Island, New York, USA}

\begin{document}

The physical pion mass is small but not zero.  In consequence, the conventional wisdom is that for two flavors of quarks there is no true thermodynamic chiral phase transition at finite temperature $T$ and zero baryon chemical potential $\mu$.  Instead, there is expected to be a curve of first-order phase transition in the $\mu$-$T$ plane that terminates in a second-order phase transition at some critical point $(\mu_c,T_c)$.  The location of the critical point obviously depends on the physical (vacuum) pion mass.  This topic has been under intense theoretical study using various effective field theory models, such as the Namu Jona-Lasinio model \cite{asakawa89,berges98,scavenius01}, a composite operator model \cite{barducci}, a random matrix model \cite{halasz98}, a linear $\sigma$ model \cite{scavenius01}, an effective potential model \cite{hatta02}, and a hadronic bootstrap model \cite{antoniou02}, as well as various implementations of lattice QCD \cite{fodor02,ejiri03,forcrand03,gavai05}.  Reviews of the subject were presented in the past few years \cite{stephanov}.  It is also of great interest because collisions between heavy nuclei at medium to high energy, such as at the future Facility for Antiproton and Ion Research (FAIR), or possible low energy runs at the Relativistic Heavy Ion Collider (RHIC), may provide experimental information on the phase diagram in the vicinity of a critical point.

In this work we study the phase diagram of the linear $\sigma$ model coupled to two flavors of identical mass quarks.  Whereas the $\sigma$ model is an oft-used effective model that represents some of the essential features of the chiral dynamics of QCD, the reason to couple the fields to quarks is less obvious.  One argument is based on the existence of the critical point itself.  If a critical point exists, then one can go around it without crossing the curve of the first-order phase transition.  The effective degrees of freedom should not change too much in following such a path. Therefore, if constituent quarks are considered to be reasonably useful degrees of freedom on the higher temperature side then they should be useful on the lower temperature side too.  For this reason we use quarks to carry the baryon number but acknowledge the resulting uncertainty in the results.  The present work extends that of Refs. \cite{scavenius01,fraser,carter97,carter00,Mocsy2004} in several ways.  We include thermal fluctuations of the meson and fermion fields, which can be important at finite temperature when the magnitude of the fluctuations becomes comparable to or greater than the mean values.  We scan vacuum pion masses from zero to over 300 MeV, which may be particularly useful for comparison with lattice gauge theory with different quark masses.  Even when all other parameters of the model are fixed, we find an interesting phase diagram that may have two, one or no critical points depending on the value of the vacuum pion mass.  

The Lagrangian is
\begin{equation}
{\cal L} = \frac{1}{2}\left(\partial_{\mu}\boldsymbol{\pi}\right)^2
+ \frac{1}{2}\left(\partial_{\mu}\sigma\right)^2 - 
U(\sigma,\boldsymbol{\pi}) 
+ \bar{\psi} \left[ i \! \not\!\partial -
g \left( \sigma + i \gamma_5
\boldsymbol{\tau} \cdot \boldsymbol{\pi} \right) \right] \psi \, ,
\end{equation}
where
\begin{equation}
U(\sigma,\boldsymbol{\pi}) = \frac{\lambda}{4}\left( \sigma^2+ \boldsymbol{\pi}^2 - f^2 \right)^2 - H \sigma 
\end{equation}
in an obvious notation.  The SU$(2)_L\times$SU$(2)_R$ chiral symmetry is explicitly broken by the term $H \sigma$, which gives the pion a mass.  The parameters in the Lagrangian are constrained by fixing the pion decay constant $f_{\pi} = 92.4$ MeV, the $\sigma$ mass $m_{\sigma} = 700$ MeV, and the quark mass is set to one-third of the nucleon mass $m_q = 313$ MeV.  The vacuum pion mass $m_{\pi}$ is varied from 0 to $m_{\sigma}/2 = 350$ MeV.  To create an effective mesonic model, we integrate out the quark degrees of freedom in the usual way \cite{kap}, such that
\begin{equation}
\ln {\cal Z}_{\rm quark}=\ln \det {\it D} \, ,
\end{equation}
where ${\it D}$ is the inverse quark propagator.  This leads to an effective Lagrangian
\begin{equation}
{\cal L}_{\rm eff} = \frac{1}{2}\left(\partial_{\mu}\boldsymbol{\pi}\right)^2
+ \frac{1}{2}\left(\partial_{\mu}\sigma\right)^2 -
U_{\rm eff}(\sigma,\boldsymbol{\pi}) \, ,
\label{effL} 
\end{equation}
where
\begin{equation}
U_{\rm eff}(\sigma,\boldsymbol{\pi}) =
U(\sigma,\boldsymbol{\pi})
- \frac{T}{V} \, \ln {\cal Z}_{\rm quark}(\sigma,\boldsymbol{\pi})
\label{effU} 
\end{equation}
is the effective potential.  We decompose the scalar field into a condensate $v$ plus a fluctuation $\Delta$.  The thermodynamic potential is computed from
\begin{equation}
\Omega=\langle U_{\rm eff}\rangle-\frac{1}{2}m_\sigma^2\langle\Delta^2\rangle
-\frac{1}{2}m_\pi^2\langle\boldsymbol{\pi}^2\rangle+\Omega_\sigma+\Omega_\pi \, ,
\end{equation}
where $\Omega_\sigma$ and $\Omega_\pi$ are the independent particle contributions from the $\sigma$ and pion quasiparticles.  Angular brackets refer to thermal averaging.  The condensate is determined self-consistently by
\begin{equation}
\left\langle\frac{\partial U_{\rm eff}}{\partial v}\right\rangle=0 
\end{equation}
while the quasiparticle masses are determined self-consistently by
\begin{equation}
m_\sigma^2 = \left\langle
\frac{\partial^2 U_{\rm eff}}{\partial\Delta^2} \right\rangle \;\;\;\;\;\;\;
m_\pi^2 = \left\langle
\frac{\partial^2 U_{\rm eff}}{\partial\pi_i^2}\right\rangle \, .
\end{equation}
The thermal fluctuations are determined by
\begin{equation}
\langle\Delta^2\rangle=2\frac{\partial\Omega_\sigma}{\partial m_\sigma^2}
=\frac{1}{2\pi^2}\int\limits_0^\infty dp\frac{p^2}{E_\sigma}
\frac{1}{e^{\beta E_\sigma}-1}
\end{equation}
and
\begin{equation}
\langle\boldsymbol{\pi}^2\rangle=2\frac{\partial\Omega_\pi}{\partial m_\pi^2} = \frac{3}{2\pi^2}\int\limits_0^\infty dp\frac{p^2}{E_\pi}\frac{1}{e^{\beta E_\pi}-1} \, .
\end{equation}
The techniques for evaluating the thermal average $\langle U_{\rm eff} \rangle$ and its derivatives are nontrivial and were developed in Refs. \cite{fraser,carter97,carter00}. The resulting equations must be solved (very accurately) using numerical methods.  All thermodynamic identities were verified both analytically and numerically.
  
An example of the pressure versus chemical potential at fixed temperature is shown in the left panel of Fig. 1.  For the higher temperature of 80 MeV there is only one self-consistent solution.  For the lower temperature of 50 MeV, there is a unique solution at large $\mu$ and another unique solution at small $\mu$.  For a range of $\mu$ centered about 900 MeV there are three solutions: One is associated with a continuation of the low-density phase, one is associated with a continuation with the high-density phase, and the third solution (not shown in the figure) is an unstable phase.  The point where the two curves cross is the location of the phase transition.  In this example it is first-order since there is a discontinuity in the slope $\partial P(\mu,T)/\partial \mu \equiv n_B$.  Where each curve terminates is the limit of metastability for that phase.  The thermodynamically favored phase is the one with the largest pressure.  At some temperature between 50 and 80 MeV the slopes are equal at the crossing point, there are no metastable phases, and the second derivative is discontinuous.  This is indicative of a second-order phase transition.  The location of this point in the $\mu$-$T$ plane is the critical point.

\begin{figure}[ht]
\centering
\includegraphics[width=0.49\textwidth]{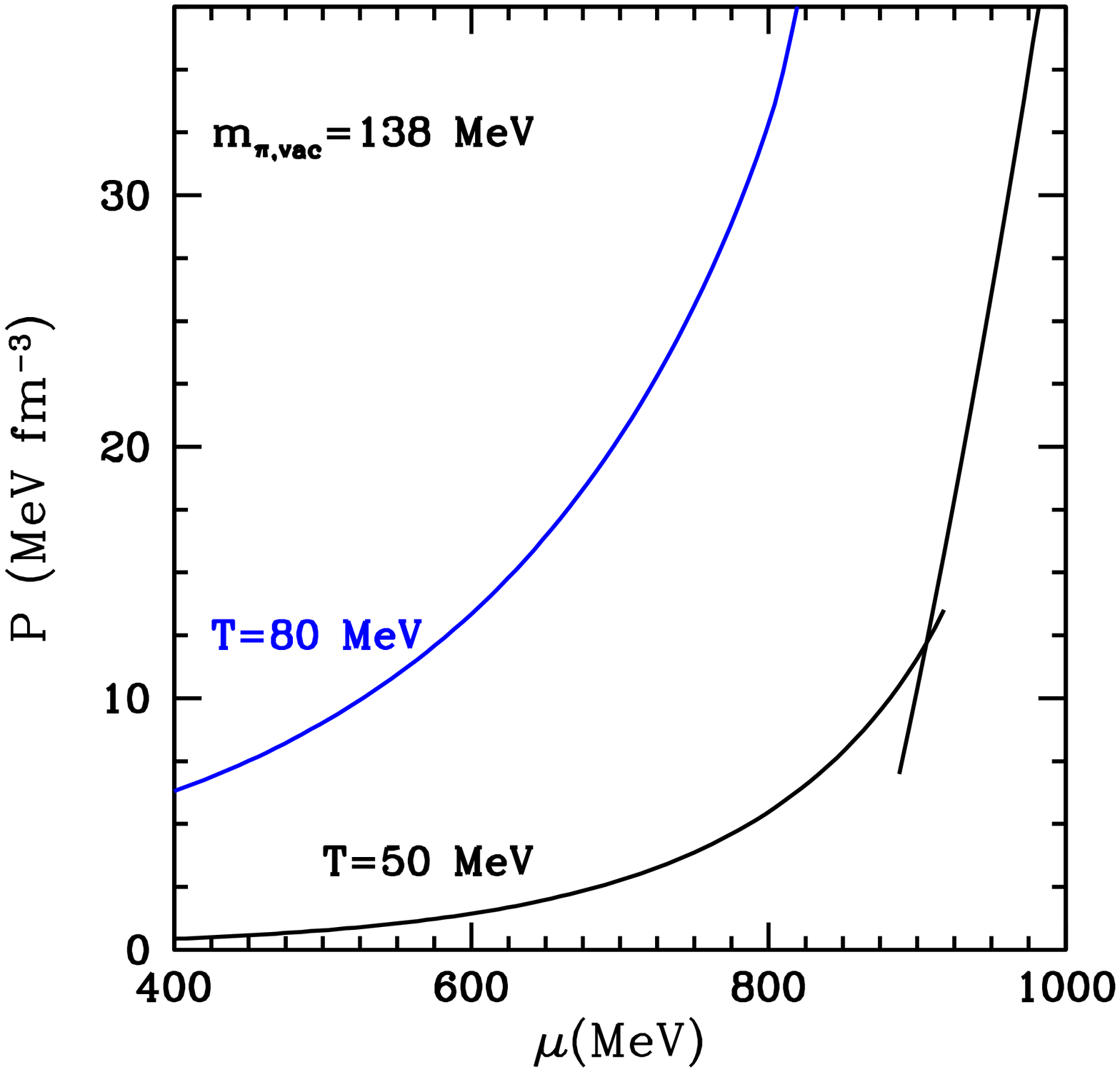}
\includegraphics[width=0.49\textwidth]{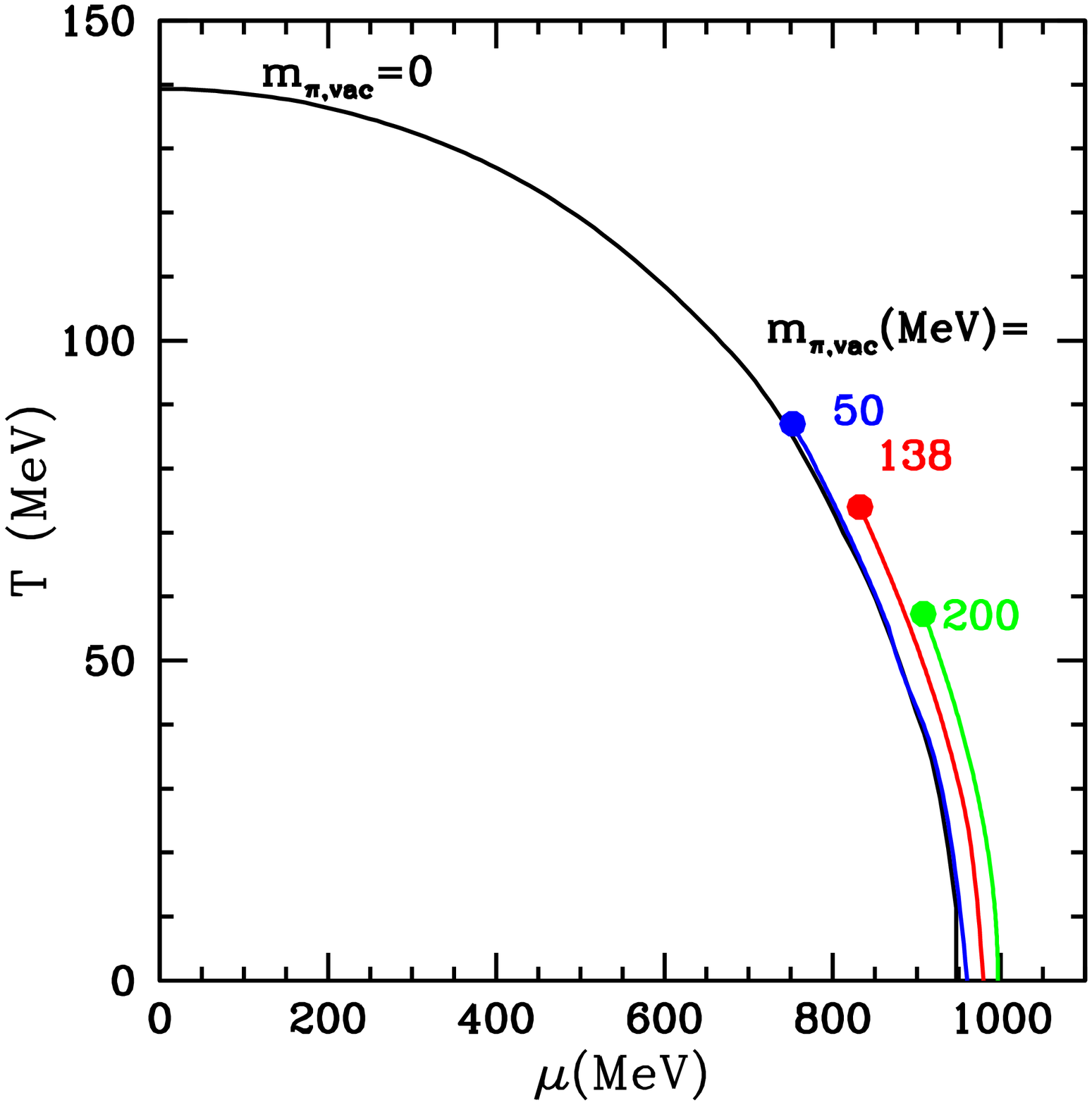}
\caption[]{Left panel: Pressure versus baryon chemical potential for temperatures below and above $T_c$ for the physical vacuum pion mass.  Right panel: Phase diagram for illustrative values of the vacuum pion mass.}
\label{Fig1}
\end{figure}

The phase diagram for a sampling of vacuum pion masses is shown in the right panel of Fig. 1.  Generically there is a curve of first-order phase transition starting on the $\mu$ axis and arching to the left.  This curve terminates at a critical point. For $m_{\pi,{\rm vac}} = 321$ MeV the critical point sits on the $T = 0$ axis, and for $m_{\pi,{\rm vac}} > 321$ MeV there is no phase transition at all.  Of course the precise numbers depend on the constants in this model, such as the vacuum $\sigma$ mass and what value one assigns to the coupling of the quark field to the $\sigma$ field, but the results are in line with expectations \cite{stephanov}.

Something very interesting happens as the vacuum pion mass is decreased from 50 MeV to 0.  The left panel of Fig. 2 shows the phase diagram for a vacuum pion mass of 35 MeV.  There are now two critical points!  There is a line of first-order phase transition beginning on the $\mu$ axis and arching to the left to end at a critical point of $\mu_{c1} \approx 725$ MeV and $T_{c1} \approx 92$ MeV and there is another line of first-order phase transition beginning on the $T$ axis and arching to the right to end at a critical point of $\mu_{c2} \approx 240$ MeV and $T_{c2} \approx 137$ MeV.  The latent heat curves are shown in the right panel of Fig. 2.  In the vicinity of $\mu = 575$ MeV and $T = 110$ MeV the latent heat gets pinched to zero for some critical value of the vacuum pion mass between 0 and 35 MeV.

\begin{figure}[ht]
\centering
\includegraphics[width=0.49\textwidth]{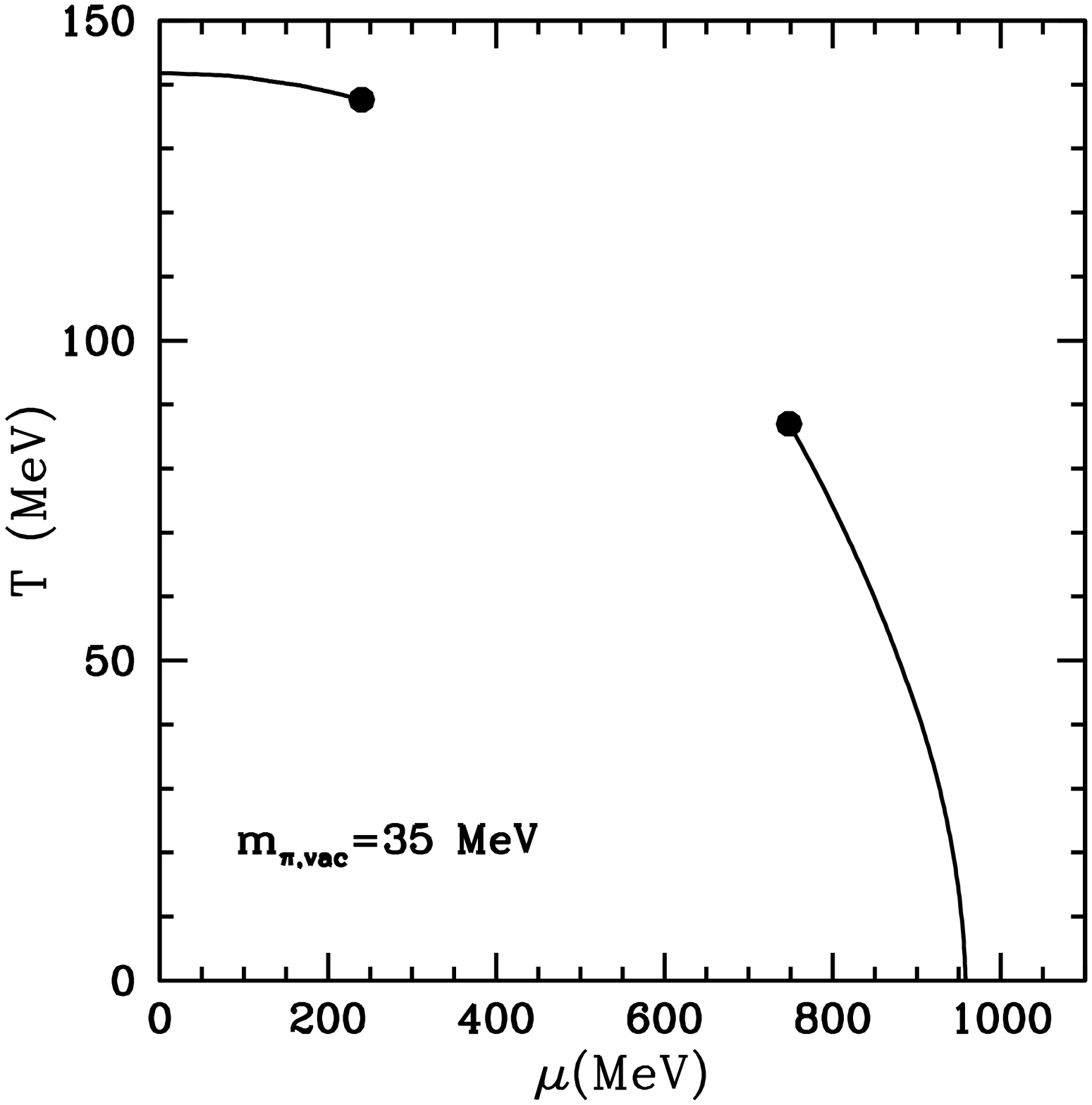}
\includegraphics[width=0.49\textwidth]{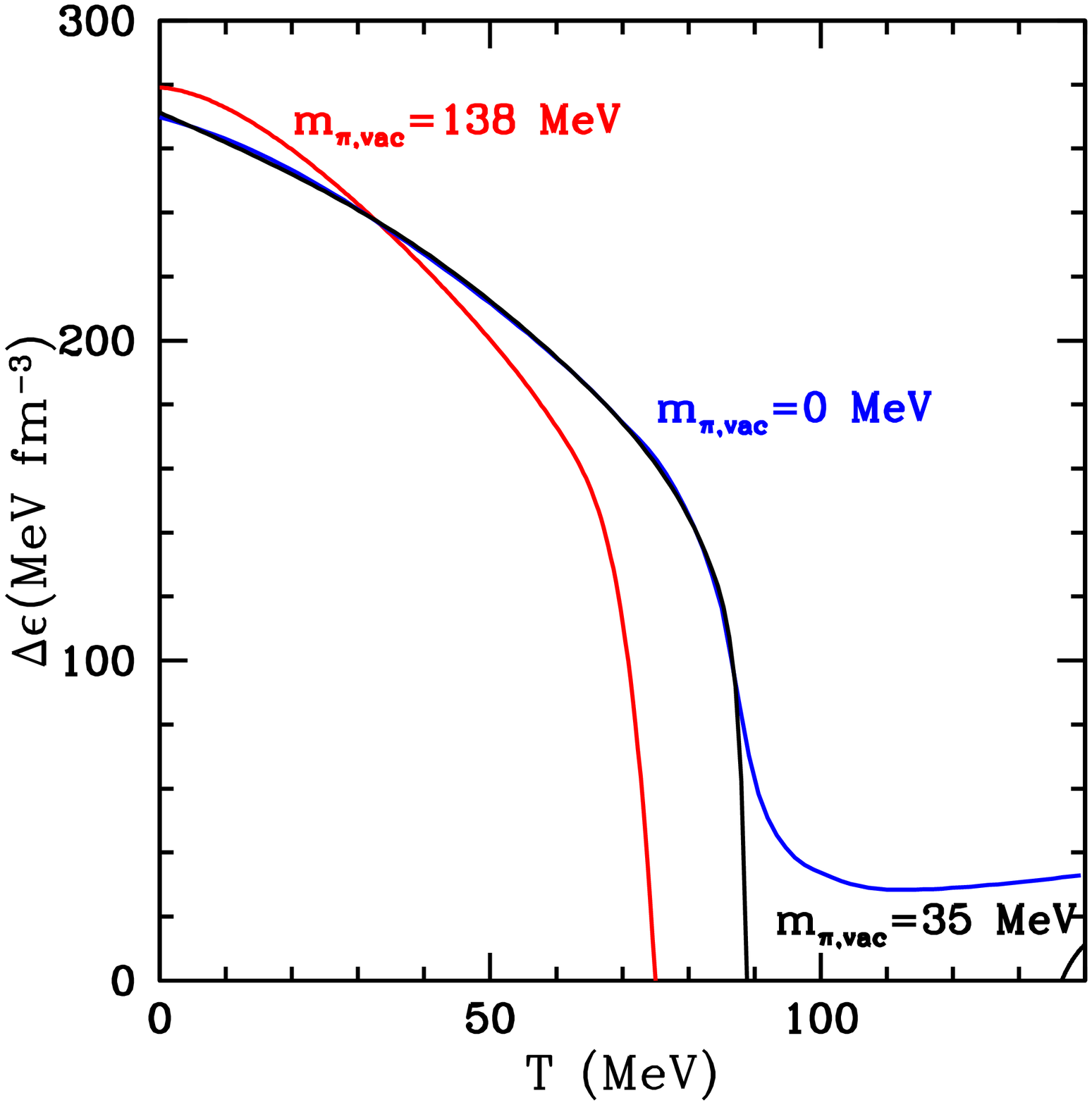}
\caption[]{Left panel: Phase diagram for a vacuum pion mass of 35 MeV.  There are two critical points.  Right panel: Latent heat along the coexistence curve for illustrative values of the vacuum pion mass.  Note the second region in the lower right corner for the 35 MeV mass.}
\label{Fig2}
\end{figure}

The results shown in the right panel of Fig. 1 illustrate the conventional view on the nature and location of the critical point.  The result shown in the left panel of Fig. 2 is unconventional or exotic.  To the best of our knowledge, this is the first time a model calculation has resulted in such a phase diagram.  
It is well known that the nature of the phase transition or crossover for two flavors of quarks is sensitive to such details as the strength of the axial U(1) anomaly and the value of the vacuum $\sigma$ mass \cite{Pisarski,Lenaghan,Chandra}.  Even for two flavors of massless quarks with 
$\mu = 0$ the order of the transition is not known.  In particular, the authors of Ref. \cite{Giacomo} found that their Monte Carlo results were substantially consistent with a first-order transition with respect to scaling of the specific heat and the chiral condensate, but not with respect to the chiral susceptibility.  It cannot be expected that a model as simple as the one studied here can make definitive predictions for what actually happens in QCD.  However, it can serve to illustrate the possibilities.  One such possibility is sketched in Fig. 3.  There are critical surfaces which delineate regions of first order phase transitions.  Some lattice calculations \cite{forcrand03} show the surface bending towards the $\mu$-axis.  This cannot continue indefinitely, and it is conceivable that the surface will eventually curve away from it.  This would indicate two critical points for certain ranges of the quark masses.  

\begin{figure}[bh]
\centering
\includegraphics[width=0.71\textwidth]{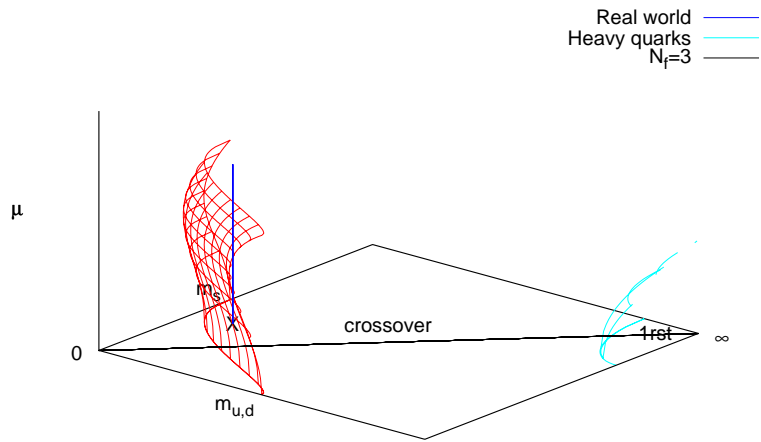}
\caption[]{Sketch of what might happen with three flavors of quarks.  Due to a change in the curvature of the surface more than one critical point might arise.  Figure courtesy of Ph. de Forcrand.}
\label{Fig3}
\end{figure}

\newpage

Further work with this and related models can be easily identified.  Changing the other parameters in the model, such as the vacuum $\sigma$ mass or the strength of the coupling of the quarks to the chiral fields, will obviously change the particular values of the pion masses for which two, one or no critical points exist.  Recent studies in the Nambu Jona-Lasinio model with varying vector coupling strength show analogous critical behavior \cite{Kenji}.  
It would be interesting to extend the $\sigma$ model to include strangeness and to vary both the vacuum pion and kaon mass using the self-consistent techniques of the present paper.  It should be straightforward to calculate shear and bulk viscosities and thermal conductivity along the lines of Refs. \cite{Paech,Sasaki}.  These transport coefficients could be used in fluid dynamic modeling of heavy-ion collisions at FAIR and for low-energy runs at RHIC.  More details on this work have been, or will be, presented elsewhere \cite{KapBow,BowmanPhD}.  

\section*{Acknowledgments}
We are grateful to E. E. Kolomeitsev, Ph. de Forcrand, O. Philipsen, and M. Stephanov for discussions.  This work was supported by the U.S. Department of Energy (DOE) under Grant No. DE-FG02-87ER40328.


\begin{thebibliography}{99}

\bibitem{asakawa89} M. Asakawa and K. Yazaki, Nucl. Phys. {\bf A504}, 668 (1989).

\bibitem{berges98} J. Berges and K. Rajagopal, Nucl. Phys. {\bf B538}, 215 (1999).

\bibitem{scavenius01}
O. Scavenius, A. M\`ocsy, I. N. Mishustin, and D. H. Rischke, Phys. Rev. C 
{\bf 64}, 045202 (2001).

\bibitem{barducci} A. Barducci, R. Casalbuoni, S. De Curtis, R. Gatto, and G. Pettini, Phys. Lett. {\bf B231}, 463 (1989); Phys. Rev. D {\bf 41}, 1610 (1990); A. Barducci, R. Casalbuoni, G. Pettini, and R. Gatto, {\it ibid.} 
{\bf 49}, 426 (1994).

\bibitem{halasz98} M. A. Halasz, A. D. Jackson, R. E. Shrock, M. A. Stephanov, and J. M. Verbaarschot, Phys. Rev. D {\bf 58}, 096007 (1998).

\bibitem{hatta02} Y. Hatta and T. Ikeda, Phys. Rev. D {\bf 67}, 014028 (2003).

\bibitem{antoniou02} N. G. Antoniou and A. S. Kapoyannis, Phys. Lett. 
{\bf B563}, 165 (2003).

\bibitem{fodor02} Z. Fodor and S. D. Katz, J. High Energy Phys. 03 (2002) 014; {\it ibid.} 04 (2004) 050.

\bibitem{ejiri03} S. Ejiri, C. R. Allton, S. J. Hands, O. Kaczmarek, F. Karsch, E. Laermann, and C. Schmidt, Prog. Theor. Phys. Suppl. {\bf 153}, 118 (2004).

\bibitem{forcrand03}
Ph. de Forcrand and O. Philipsen, Nucl. Phys. {\bf B642}, 290 (2002);
{\bf B673}, 170 (2003);  Nucl. Phys. Proc. Suppl. {\bf 129}, 521 (2004); J. High Energy Phys. 11 (2008) 012.

\bibitem{gavai05} R. V. Gavai and S. Gupta, Phys. Rev. D {\bf 71}, 114014 (2005).

\bibitem{stephanov}
M. Stephanov, Prog. Theor. Phys. Suppl. {\bf 153}, 139 (2004);
Int. J. Mod. Phys. A {\bf 20}, 4387 (2005); PoS(LAT2006)024.

\bibitem{fraser} 
C. M. Fraser, {\it{Z. Phys.}} C {\bf28}, 101 (1985); I. J. R. Aitchison and C. M. Fraser, Phys. Rev. D {\bf31}, 2605 (1985).

\bibitem{carter97} 
G. W. Carter, P. J. Ellis, and S. Rudaz, Nucl. Phys. {\bf A618}, 317 (1997).

\bibitem{carter00} 
G. W. Carter, O. Scavenius, I. N. Mishustin, and P. J. Ellis, Phys. Rev. C {\bf61},045206 (2000).

\bibitem{Mocsy2004}
A. M\`ocsy, I. N. Mishustin and P. J. Ellis, Phys. Rev. C {\bf 70}, 015204 (2004).

\bibitem{kap} 
J. I. Kapusta and C. Gale, {\it Finite-Temperature Field Theory} (Cambridge University Press, Cambridge, UK, 2006).

\bibitem{Pisarski}
R. D. Pisarski and F. Wilczek, Phys. Rev. D {\bf 29}, 338 (1984).

\bibitem{Lenaghan}
J. T. Lenaghan, Phys. Rev. D {\bf 63}, 037901 (2001).

\bibitem{Chandra}
S. Chandrasekharan and A. C. Mehta, Phys. Rev. Lett. {\bf 99}, 142004 (2007).

\bibitem{Giacomo} G. Cossu, M. D'Elia, A. Di Giacomo, and C. Pica, 
PoS(LAT2007)219.

\bibitem{Kenji}
K. Fukushima, Phys. Rev. D {\bf 78}, 114019 (2008).

\bibitem{Paech}
K. Paech and S. Pratt, Phys. Rev. C {\bf 74}, 014901 (2006).

\bibitem{Sasaki}
C. Sasaki and K. Redlich, Phys. Rev. C {\bf 79}, 055207 (2009).

\bibitem{KapBow} E. S. Bowman and J. I. Kapusta, Phys. Rev. C {\bf 79}, 015202 (2009).

\bibitem{BowmanPhD}
E. S. Bowman, Ph.D. thesis, University of Minnesota, 2009, and manuscript in preparation.


\end{thebibliography}
\end{document}